# On the optical Stark effect of excitons in InGaAs prolate ellipsoidal quantum dots

Le Thi Ngoc Bao[1*], Duong Dinh Phuoc[2], Le Thi Dieu Hien[1,2], and Dinh Nhu Thao[2**]

[1] *Hue University of Sciences, Hue University, 77 Nguyen Hue Street, Hue City, Vietnam*
[2] *Hue University of Education, Hue University, 34 Le Loi Street, Hue City, Vietnam*

In this paper, we study the exciton absorption spectra in InGaAs prolate ellipsoidal quantum dots when a strong pump laser resonant with electron quantized levels is active. Our obtained results by renormalized wavefunction theory show that, under suitable conditions, the initial exciton absorption peak is split into two new peaks as the evidence of the existence of the three-level optical Stark effect of excitons. We have suggested an explanation of the origin of the effect as well as investigating the effect of pump field energy, size, and geometric shape of the quantum dots on effect characteristics. The comparison with the results obtained in the spherical quantum dots implies the important role of geometric shape of the quantum structures when we examine this effect.

## 1. Introduction

One of the low dimensional structures recently attracting much interest from researchers is the semiconductor quantum dots [1-4], which are structures that confine carriers in all three spatial dimensions. Quantum dot structures have many interesting applications, such as producing artificial atoms and molecules [5], single-electron transistors [6], and quantum dot lasers [7]. In recent years, there has seen many studies on optical properties in quantum dots with different shapes, for example cylindrical, cubic, and spherical quantum dots [8-12]. These studies show that optical properties of quantum dots are highly dependent on external fields and size of quantum dots. It is worth mentioning that the shape of quantum dots also makes a huge difference to the optical properties of quantum dots [13]. Therefore, quantum dots are expected to create even more breakthrough applications in the future.

The optical Stark effect of excitons is one of the unique optical properties of bulk materials in general, and low dimensional structures in particular. The optical Stark effect of excitons occurs due to the coupling of the two exciton states under the excitation of near resonant beam [14]. Scientists divided this effect into two types as follows. First, two-level optical Stark effect arises when a pump laser of strong intensity couples exciton ground state and an exciton excited state in the quantum system [14,15]. Second, three-level optical Stark effect is the result of a couple of two excited states of exciton under the control of a pump laser beam of lower intensity [16,17]. The latter has received more attention from researchers because they are more likely to occur and have better potential application. In general, optical Stark effect has potential applications in manufacturing ultrafast optical switching of future optical devices [11,18], optical modulators [19], mesoporous hybrid multifunctional system [20], and optically controlled field-effect transistors [21]. The optical Stark effect of excitons in quantum dots has also been of interest in both experimental and theoretical research [11,22-24], but mainly for spherical quantum dots. The investigation of this effect in a more complicated and more practical shaped quantum dot structures, such as ellipsoidal quantum dots, has not been carried out in detail.

In this paper, we study theoretically the three-level optical Stark effect of excitons in InGaAs prolate ellipsoidal quantum dots. We applied the renormalized wavefunction method to investigate the dependence of the exciton absorption spectra on the external fields, size, and shape of quantum dot, when there exists the optical Stark effect of excitons. In addition, the effects of the pump laser energy, size, and geometric shape of quantum dots on the behavior of the effect are also clarified. The article includes the following main sections: part 2 presents model and theory, part 3 presents the main results and discussion, and conclusions are presented in part 4.

## 2. Model and Theory

*2.1. Wavefunctions and energy levels of electron and hole in prolate ellipsoidal quantum dots*

In this section, we present the wavefunctions and energy levels of electron and hole in prolate ellipsoidal quantum dots. We consider a prolate ellipsoidal quantum dot with rotational symmetry around the *z* axis, with *a* and *c* are its semi-axes along the xOy plane and stay in the z-direction, respectively; in which *x*, *y*, *z* are the coordinates in Cartesian coordinate system with its origin at the ellipsoidal symmetry center (Fig. 1). Simply put, we assume that prolate ellipsoidal quantum dot is set under the effect of an infnite potential of the form [25-29]

$$U(\vec{r}) = \begin{cases} 0, & \text{where} \quad 0 < S(\vec{r}_i) < 1 \\ \infty, & \text{where} \quad S(\vec{r}_i) \geq 1 \end{cases}, \quad (1)$$

where $S(\vec{r}_i)$ depends on parameters *a* and *c* which are semi-axes of the ellipsoidal quantum dot, as follows

$$S(\vec{r}_i) = \frac{x^2 + y^2}{a^2} + \frac{z^2}{c^2}, \quad (2)$$

with $c > a$, and the surface of the prolate ellipsoidal quantum dot has the shape as in Fig. 1.

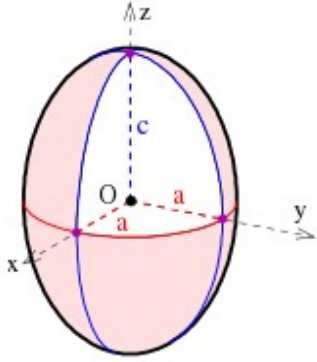

Fig. 1. Illustration of a prolate ellipsoidal quantum dot [30].

To conveniently examine the prolate ellipsoidal quantum dots, in this study, we use the prolate spheroidal coordinates

$$\begin{cases} x = f\sqrt{(\xi^2-1)(1-\eta^2)}\cos\varphi \\ y = f\sqrt{(\xi^2-1)(1-\eta^2)}\sin\varphi, \\ z = f\xi\eta \end{cases} \quad (3)$$

where $1 \leq \xi < +\infty$, $-1 \leq \eta \leq +1$, and $f$ is the parameter. The envelope wavefunctions of electron (hole) in prolate ellipsoidal quantum dot then has the following form [25-29]

$$\Psi(\vec{r}) = \Psi_{nlm}^{e,h}(\xi,\eta,\varphi) = A_{nlm} J_{lm}^{(1)}(h,\xi) S_{lm}^{(1)}(h,\eta) e^{im\varphi}, \quad (4)$$

where $n = 1, 2, 3, ...$; $l = 0, 1, 2, 3, ...$; $m = -l, ..., 0, ..., +l$; $J_{lm}^{(1)}(h,\xi)$ and $S_{lm}^{(1)}(h,\eta)$ are prolate radial and prolate angular spheroidal functions of the first kind, respectively; and $A_{nlm}$ is the normalization coefficient

$$A_{nlm} = \sqrt{\frac{\chi^3}{2\pi c^3 e^3 \int_1^{\bar{\xi}} \int_{-1}^{+1} (\xi^2+\eta^2) J_{lm}^{(1)*}(h,\xi) S_{lm}^{(1)*}(h,\eta) J_{lm}^{(1)}(h,\xi) S_{lm}^{(1)}(h,\eta) d\xi d\eta}}, \quad (5)$$

where the ellipsoidal aspect ratio is defined as

$$\chi = \frac{c}{a}, \quad (6)$$

and $e$ is the ellipsoidal eccentricity

$$e = \sqrt{1 - \frac{1}{\chi^2}}. \quad (7)$$

The quantized energy levels of electron (hole) are determined as follows

$$\varepsilon_{nlm}^{e,h} = \frac{\hbar^2 k_{nlm}}{2m_{e,h}^*}, \quad (8)$$

where $m_{e,h}^*$ is the effective mass of electron (hole) and

$$h = f\sqrt{k_{nlm}}. \quad (9)$$

The values of $h$ are determined from the boundary condition

$$J_{lm}^{(1)}(h,\bar{\xi}) = 0, \quad (10)$$

where

$$\bar{\xi} = \frac{1}{\sqrt{1-\frac{1}{\chi^2}}} = \frac{1}{e}, \qquad (11)$$

and

$$f = \frac{c}{\bar{\xi}} = c \cdot e. \qquad (12)$$

The values of the parameter $h$ depend on the values of the indices $n$, $l$, and $m$. When $h \to 0$ (or $f \to 0$), the prolate ellipsoidal quantum dot becomes the spherical quantum dot; therefore, $\chi \to 1$. In that condition, the envelope wavefunctions of electron (hole) in quantum dot looks like the following form [31]

$$\Psi_{nlm}^{S(e,h)}(r,\theta,\varphi) = \sqrt{\frac{2}{R^3}} \frac{j_l\left(\chi_{nl}\frac{r}{R}\right)}{j_{l+1}(\chi_{nl})} Y_{lm}(\theta,\varphi), \qquad (13)$$

where $Y_{lm}(\theta,\varphi)$ is the spherical harmonic function; $j_l(r)$ is the spherical Bessel function of order $l$ ($l$ is a non-negative integer, $l = 0,1,2...$) and $\chi_{nl}$ is its $n$th zero. The expression of the quantized energy levels of electron (hole) corresponding to wavefunctions (13) is

$$E_{nl}^{S(e,h)} = \frac{\hbar^2 \chi_{nl}^2}{2m_{e,h}^* R^2}. \qquad (14)$$

In the Eqs. (13) and (14), the indices $n$, $l$, and $m$ are principle, orbital, and azimuthal quantum numbers, respectively. However, for ellipsoidal quantum dots, since its spherical symmetry no longer exists, so the index $l$ in the wavefunction and energy expressions of the particle in equations (4) and (8) no longer represent the orbital quantum numbers. However, in this paper, we still use the set of indices $n$, $l$, $m$ in Eqs. (4) and (8) in order to have a one-to-one correspondence between prolate ellipsoidal and spherical quantum dot when $\chi \to 1$ [26]. The volume of prolate ellipsoidal quantum dot of semi-axes $a$ and $c$ is determined as follows [32]

$$V = \frac{4}{3}\pi a^2 c = \frac{4}{3}\pi a^3 \chi = \frac{4}{3}\pi R_S^3, \tag{15}$$

where $R_s = a\sqrt[3]{\chi}$ is the radius of a sphere with the same volume.

In the effective mass approximation and envelope-function theory, the total wavefunction of electrons (holes) in a prolate ellipsoidal quantum dot with an infinite potential has the following form

$$\Lambda_{nlm}^{e,h}(\vec{r}) = u_{c,v}(\vec{r})\Psi_{nlm}^{e,h}(\vec{r}), \tag{16}$$

where $\vec{r} = (\xi, \eta, \varphi)$ and $u_{c,v}(\vec{r})$ are the periodic Bloch functions in conduction and valence band.

As we set the zero energy at the top of the valence band, the expression of the quantized energy levels of electron and hole is then determined as follows

$$E_{nlm}^{e} = E_g + \frac{\hbar^2 k_{nlm}}{2m_e^*}, \tag{17}$$

and

$$E_{nlm}^{h} = \frac{\hbar^2 k_{nlm}}{2m_h^*}, \tag{18}$$

where $E_g$ is the bandgap of the semiconductor.

*2.2. Intersubband optical transition*

In this paper, we study the three-level optical Stark effect of excitons in prolate ellipsoidal quantum dots. Thus, we need to examine a three-level system of exciton and thus result in a three-level diagram of electron and hole, which consist of the lowest level $E_{100}^{h}$ corresponding to the first quantized state of the hole in the valence band $|0\rangle$, and the other two levels have the energy of $E_{100}^{e}$ and $E_{110}^{e}$ corresponding to two lowest quantized states of the electrons in the conduction band $|1\rangle$ and $|2\rangle$ as illustrated in Fig. 2.

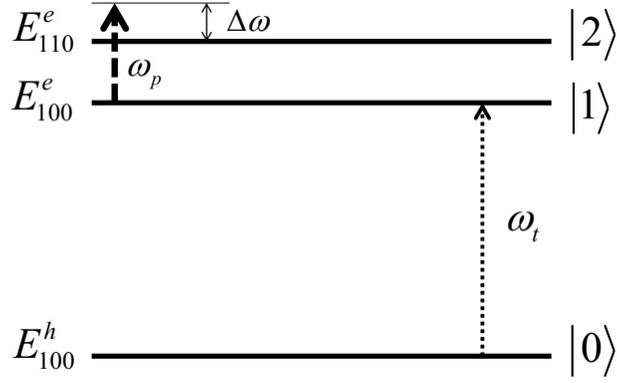

Fig. 2. Three-level energy diagram of electron and hole. $E_{100}^{h}$ is the lowest energy level of the hole corresponding to state $|0\rangle$; $E_{100}^{e}$ và $E_{110}^{e}$ are the two lowest energy levels of electrons corresponding to states $|1\rangle$ và $|2\rangle$; $\omega_p$ and $\omega_t$ are frequency of the pump or probe laser, respectively; $\Delta\omega$ is pump laser detuning.

When the electromagnetic field is not so strong, we can omit higher-order term and using some gauges and approximations, the electron-electromagnetic field interaction Hamiltonian then can be written as follows [31,33]

$$\hat{H}_{\text{int}} = -\frac{q}{m_0} \frac{A_x e^{-i\omega_x t}}{i\omega_x} \vec{n} \cdot \hat{\vec{p}} \equiv \hat{V}_x e^{-i\omega_x t}, \qquad (19)$$

in which we set

$$\hat{V}_x = -\frac{q}{m_0} \frac{A_x}{i\omega_x} \vec{n} \cdot \hat{\vec{p}}, \qquad (20)$$

where $q$, $m_0$, and $\vec{p}$ are the bare mass, the charge, and the momentum of the particle, respectively; $A_x$ and $\omega_x$ are respectively magnitude and frequency of the laser, with $x$ clarifying if that is a pump or probe laser; and $\vec{n}$ is the unit vector pointing to the wave propagation direction.

In order to investigate the three-level optical Stark effect of excitons in prolate ellipsoidal quantum dots, we need to examine two optical transitions, which are the intersubband transition between electron levels under a strong pump laser and the interband transition between two lowest levels of hole and electron under a probe laser. In this section, we examine the matrix element for the optical intersubband transition between electron levels $E_{100}^e$ and $E_{110}^e$ under the effect of pump laser. The two quantized states of electrons are defined as follows

$$\begin{cases} |1\rangle \equiv \Lambda_{100}^e(\vec{r}) = u_c(\vec{r}) \Psi_{100}^e(\vec{r}) \\ |2\rangle \equiv \Lambda_{110}^e(\vec{r}) = u_c(\vec{r}) \Psi_{110}^e(\vec{r}) \end{cases}. \tag{21}$$

The matrix element for an intersubband transition between two electron quantized levels defined as follows

$$v_{21} = \langle 2 | \hat{H}_{int} | 1 \rangle = \langle 2 | \hat{V}_p | 1 \rangle e^{-i\omega_p t} \equiv V_{21} e^{-i\omega_p t}, \tag{22}$$

in which

$$V_{21} = -\frac{qA_p}{m_0 i \omega_p} \langle 2 | \hat{V} | 1 \rangle = -\frac{qA_p}{m_0 i \omega_p} \langle \Psi_{110}^e(\vec{r}) | \vec{n}\hat{p} | \Psi_{100}^e(\vec{r}) \rangle, \tag{23}$$

where $A_p$ and $\omega_p$ are magnitude and frequency of the pump laser. Equation (23) can be rewritten as follows

$$V_{21} = \frac{qA_p}{m_0 i \omega_p} \frac{m_e^*}{i\hbar} \left( E_{110}^e - E_{100}^e \right) \langle \Psi_{110}^e(\vec{r}) | \vec{n}\vec{r} | \Psi_{100}^e(\vec{r}) \rangle. \tag{24}$$

From Eq. (4), we have

$$\begin{cases} \Psi_{110}^e(\vec{r}) = A_{110} J_{10}^{(1)}(h,\xi) S_{10}^{(1)}(h,\eta) e^{i0\varphi} = A_{110} J_{10}^{(1)}(h,\xi) S_{10}^{(1)}(h,\eta) \\ \Psi_{100}^e(\vec{r}) = A_{100} J_{00}^{(1)}(h,\xi) S_{00}^{(1)}(h,\eta) e^{i0\varphi} = A_{100} J_{00}^{(1)}(h,\xi) S_{00}^{(1)}(h,\eta) \end{cases}. \tag{25}$$

Choosing polarization vector along the z axis, while replacing Eq. (25) into Eq. (24), and performing some additional calculations, we have the expression of $V_{21}$ as follows

$$V_{21} = \frac{qA_p}{m_0 i\omega_p} \frac{m_e^*}{i\hbar} A_{110}^* A_{100} \left(E_{110}^e - E_{100}^e\right) 2\pi f^4 \times$$
$$\times \int_1^{\bar{\xi}} \int_{-1}^{+1} \xi\eta\left(\xi^2 - \eta^2\right) J_{10}^{(1)*}(h,\xi) S_{10}^{(1)*}(h,\eta) J_{00}^{(1)}(h,\xi) S_{00}^{(1)}(h,\eta) d\xi d\eta \quad (26)$$

From there, we have the expression of matrix element for an intersubband transition between two electron quantized levels as follows

$$v_{21} = V_{21} e^{-i\omega_p t} = \frac{qA_p e^{-i\omega_p t}}{m_0 i\omega_p} \frac{m_e^*}{i\hbar} A_{110}^* A_{100} \left(E_{110}^e - E_{100}^e\right) 2\pi f^4 \times$$
$$\times \int_1^{\bar{\xi}} \int_{-1}^{+1} \xi\eta\left(\xi^2 - \eta^2\right) J_{10}^{(1)*}(h,\xi) S_{10}^{(1)*}(h,\eta) J_{00}^{(1)}(h,\xi) S_{00}^{(1)}(h,\eta) d\xi d\eta \quad (27)$$

*2.3. The exciton absorption spectra*

*a) Without the present of the pump laser*

According to the selection rules for interband transitions in quantum dots, interband transitions only occur within the lowest energy levels of hole and electron. First, we consider the condition in which pump laser is absent, the matrix element for an interband transition of exciton only exists an allowed transition between two levels $E_{100}^h$ and $E_{100}^e$. This is an interband transition with the initial and final states described as follows

$$\begin{cases} |0\rangle = u_v(\vec{r}) \Psi_{100}^h(\vec{r}) e^{-\frac{i}{\hbar}E_{100}^h t} \\ |1\rangle = u_c(\vec{r}) \Psi_{100}^e(\vec{r}) e^{-\frac{i}{\hbar}E_{100}^e t} \end{cases} \quad (28)$$

Under the effects of probe laser, matrix element for an interband transition between two quantized levels of electron and hole is

$$T_{10} = \langle 1 | \hat{H}_{int} | 0 \rangle. \quad (29)$$

Replacing $\hat{H}_{int}$ from Eq. (19) into Eq. (29), we have the expression of matrix element for an interband transition without pump laser effect as follows

$$T_{10} = -\frac{eA_t p_{cv}}{m_0 i\omega_t} e^{\frac{i}{\hbar}\left(E^e_{100} - E^h_{100} - \hbar\omega_t\right)t}, \tag{30}$$

where $A_t$ and $\omega_t$ are magnitude and frequency of the probe laser, respectively; $p_{cv}$ is the polarization matrix element between conduction and valence band, and has the form of

$$p_{cv} = \left\langle u_c(\vec{r}) \mid \vec{n}\hat{p} \mid u_v(\vec{r}) \right\rangle. \tag{31}$$

The absorption spectra of the excitons is defined using the expression of the transition rates (or the absorption probabilities in a unit of time). According to Fermi's golden rule, the expression of transition rates is defined as [34]

$$W = \frac{2\pi}{\hbar} |T_{fi}|^2 \delta\left(E_f - E_i - \hbar\omega_t\right), \tag{32}$$

where $T_{fi}$ is the matrix element for interband transition between initial $i$ and final $f$ states; $E_i$ and $E_f$ are the corresponding energy levels of initial $i$ and final $f$ states.

From Eqs. (30) and (32), we get the expression of transition rates as follows

$$W_0 = \frac{2\pi}{\hbar} \left(\frac{qA_t p_{cv}}{m_0 \omega_t}\right)^2 \delta\left(E^e_{100} - E^h_{100} - \hbar\omega_t\right). \tag{33}$$

Applying 'Lorentz line' function [35]

$$\delta(x) = \frac{1}{\pi} \cdot \frac{\Gamma}{x^2 + \Gamma^2}, \tag{34}$$

Eq. (33) can be rewritten as follows

$$W_0 = \frac{2}{\hbar} \left(\frac{qA_t p_{cv}}{m_0 \omega_t}\right)^2 \frac{\Gamma}{\left(E^{dot}_g - \hbar\omega_t\right)^2 + \Gamma^2}, \tag{35}$$

where

$$E^{dot}_g = E^e_{100} - E^h_{100}, \tag{36}$$

and $\Gamma$ is the phenomenological linewidth of absorption peak.

*b) With the presence of the pump laser*

Under the effect of a strong pump laser resonating with the energy distance between the two quantized levels of the electron, the wavefunctions of electron are renormalized by the pump laser effect and can be described as

$$\Lambda_{mix}^{e}(\vec{r},t) = \sum_{l=0}^{1} c_{l}(t) \Lambda_{1l0}^{e}(\vec{r}) \exp(-\frac{i}{\hbar} E_{1l0}^{e} t), \quad (37)$$

where coefficients $c_l(t)$ ($l=\overline{0,1}$) are determined through the time-dependent Schrodinger equation and have the following form [12]

$$\begin{cases} c_0(t) = \dfrac{1}{2\Omega_R}\left(\alpha_1 e^{i\alpha_2 t} + \alpha_2 e^{-i\alpha_1 t}\right) \\ c_1(t) = -\dfrac{V_{21}}{2\Omega_R}\left(e^{i\alpha_1 t} - e^{-i\alpha_2 t}\right) \end{cases}, \quad (38)$$

where

$$\hbar\omega_{21} = E_{110}^{e} - E_{100}^{e}, \quad (39)$$

and

$$\begin{cases} \alpha_1 = \Omega_R - \dfrac{\Delta\omega}{2} \\ \alpha_2 = \Omega_R - \dfrac{\Delta\omega}{2} \\ \Omega_R = \left[\left(\dfrac{\Delta\omega}{2}\right)^2 + \dfrac{|V_{21}|^2}{\hbar^2}\right]^{\frac{1}{2}} \\ \Delta\omega = \omega_p - \omega_{21} \end{cases}. \quad (40)$$

Replacing coefficients $c_0(t)$ and $c_1(t)$ from Eq. (38) into Eq. (37), we get the expression of the renormalized wavefunction of electron under pump laser effect as follows

$$\Lambda_{mix}^{e}(\vec{r},t) = \dfrac{1}{2\Omega_R}\left(\alpha_1 e^{i\alpha_2 t} + \alpha_2 e^{-i\alpha_1 t}\right) e^{-\frac{i}{\hbar}E_{100}^{e}t} \Lambda_{100}^{e}(\vec{r}) - \dfrac{V_{21}}{2\hbar\Omega_R}\left(e^{i\alpha_1 t} - e^{-i\alpha_2 t}\right) e^{-\frac{i}{\hbar}E_{110}^{e}t} \Lambda_{110}^{e}(\vec{r}), \quad (41)$$

where $V_{21}$ is the quantity given in Eq. (26). Then, the Eq. (41) can be rewritten as follow

$$\Lambda^e_{mix}(\vec{r},t) = \frac{1}{2\Omega_R}\left(\alpha_1 e^{-\frac{i}{\hbar}(E^e_{100}-\hbar\alpha_2)t} + \alpha_2 e^{-\frac{i}{\hbar}(E^e_{100}+\hbar\alpha_1)t}\right)\Lambda^e_{100}(\vec{r})$$
$$- \frac{V_{21}}{2\hbar\Omega_R}\left(e^{-\frac{i}{\hbar}(E^e_{110}-\hbar\alpha_1)t} - e^{-\frac{i}{\hbar}(E^e_{110}+\hbar\alpha_2)t}\right)\Lambda^e_{110}(\vec{r}) \quad (42)$$

Therefore, the energy spectrum of electron corresponding to wavefunction given in Eq. (42) is expanded to four levels as follows

$$\begin{cases} E^{e+}_{100} = E^e_{100} + \hbar\alpha_1 \\ E^{e-}_{100} = E^e_{100} - \hbar\alpha_2 \end{cases}, \quad (43)$$

và

and

$$\begin{cases} E^{e+}_{110} = E^e_{110} + \hbar\alpha_2 \\ E^{e-}_{110} = E^e_{110} - \hbar\alpha_1 \end{cases}. \quad (44)$$

From Eqs. (43) and (44), we can rewrite the renormalized wavefunctions of the electron under the pump laser effect in Eq. (42) as follows

$$\Lambda^e_{mix}(\vec{r},t) = \frac{1}{2\Omega_R}\left(\alpha_1 e^{-\frac{i}{\hbar}E^{e-}_{100}t} + \alpha_2 e^{-\frac{i}{\hbar}E^{e+}_{100}t}\right)\Lambda^e_{100}(\vec{r}) - \frac{V_{21}}{2\hbar\Omega_R}\left(e^{-\frac{i}{\hbar}E^{e-}_{110}t} - e^{-\frac{i}{\hbar}E^{e+}_{110}t}\right)\Lambda^e_{110}(\vec{r}). \quad (45)$$

In order to defind the three-level optical Stark effect of exciton in prolate ellipsoidal quantum dots, the pump laser intensity must be significantly stronger than the probe laser. At the same time, the detuning of the pump laser to the electron levels must be much smaller than the frequency of the pump laser and band gap of active material in quantum dots

$$\Delta\omega \ll \omega_p \ll \frac{E_g}{\hbar}. \quad (46)$$

The matrix element for the interband transition between the hole state and the electron superposition state specified by the renormalized wavefunction under the effect of the probe laser when the system is irradiated by a strong resonant pump laser is defined as follows

$$T_{mix,0} = \langle \Lambda^e_{mix}(\vec{r},t) | \hat{H}_{int} | \Lambda^h_{100}(\vec{r},t) \rangle = -\frac{qA_t e^{-i\omega_t t}}{m_0 i\omega_t}\langle \Lambda^e_{mix}(\vec{r},t) | \vec{n}\hat{\vec{p}} | \Lambda^h_{100}(\vec{r},t) \rangle, \quad (47)$$

or

$$T_{mix,0} = -\frac{qA_t p_{cv}}{m_0 i \omega_t} \left[ \frac{1}{2\Omega_R} \left( \alpha_1 e^{i\alpha_2 t} + \alpha_2 e^{-i\alpha_1 t} \right) \right]^* e^{\frac{i}{\hbar}\left(E_g^{dot} - \hbar\omega_t\right)t}. \tag{48}$$

Next, we perform the similar calculation as in the previous section for the pump laser inactivity, the expression of transition rate for interband transition between the hole state and the electron superposition state when the system is irradiated by a resonant pump laser has following formula

$$W = \frac{2\pi}{\hbar} \left( \frac{eA_t p_{cv}}{m_0 \omega_t} \right)^2 \left[ \left( \frac{\alpha_1}{2\Omega_R} \right)^2 \delta\left(E_g^{dot} - \hbar\omega_t - \hbar\alpha_2\right) + \left( \frac{\alpha_2}{2\Omega_R} \right)^2 \delta\left(E_g^{dot} - \hbar\omega_t + \hbar\alpha_1\right) \right]. \tag{49}$$

Applying 'Lorentz line' function, the approximated form of the transition rate with pump laser effect can be rewritten as

$$W = \frac{2}{\hbar} \left( \frac{eA_t p_{cv}}{m_0 \omega_t} \right)^2 \left[ \left( \frac{\alpha_1}{2\Omega_R} \right)^2 \frac{\Gamma}{\left(E_g^{dot} - \hbar\omega_t - \hbar\alpha_2\right)^2 + \Gamma^2} + \left( \frac{\alpha_2}{2\Omega_R} \right)^2 \frac{\Gamma}{\left(E_g^{dot} - \hbar\omega_t + \hbar\alpha_1\right)^2 + \Gamma^2} \right]. \tag{50}$$

## 3. Results and discussion

To clarify the results obtained above, in this section, we perform the calculation for transition rate in In$_{0.53}$Ga$_{0.47}$As/In$_{0.52}$Al$_{0.48}$As prolate ellipsoidal quantum dots in two cases: without and with the pump laser. The parameters used in the calculation are as follows. The effective mass of electron and hole in the dot material In$_{0.53}$Ga$_{0.47}$As are $m_e = 0.042 m_0$ and $m_h = 0.052 m_0$; the band gap of the dot material is $E_g = 750$ meV [36]; the pump laser amplitude was chosen to be $A_p = 4 \times 10^4$ V/cm and the linewidth was chosen as the $\Gamma = 0.1$ meV. In addition, since this paper examines prolate ellipsoidal quantum dot structure. This is a low dimensional structure, so the length of the major axis $2c$ and minor axis $2a$ must be selected to be smaller the bulk exciton Bohr radius in the dot material In$_{0.53}$Ga$_{0.47}$As. Thus, we have selected the length of the semi-major axis as $a = 25$ Å to do the calculation in this section; while the length of the semi-minor axis will depend on the selected value of $\chi$ ($c = \chi \cdot a$).

In Figure 3, we depict the dependence of transition rate on the photon energy of the probe laser in a prolate ellipsoidal quantum dot with the value of the ellipsoidal aspect ratio of $\chi = 3$ in two conditions: without the pump laser effect (dashed line) and with the pump laser effect (solid line). From the figure we can see that, without the pump laser, the absorption spectra of excitons includes only one absorption peak. However, when the system is irradiated by a strong pump laser exactly resonant with the energy distance between the two quantized levels of the electron, the graph appeared two distinct equal peaks in the exciton absorption spectra. These two vertices are bilaterally symmetrical with respect to the original peak when the pump laser does not operate. Moreover, we found that the exciton absorption intensity is drastically reduced in the presence of a laser field. This reduction is the consequence of the conservation of the electronic transition rate as discussed in Ref. [11]. In short, the above results demonstrate the existence of the three-level optical Stark effect of exciton in the prolate ellipsoidal quantum dots. In addition, comparing the results found in the prolate ellipsoidal and spherical quantum dots [11], we find that they are similar. The reason is given to the fact that both prolate ellipsoidal and spherical quantum dots are quasi-zero-dimensional systems.

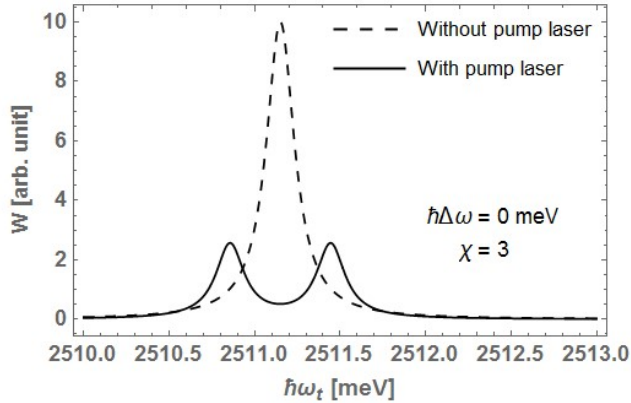

Figure 3. The dependence of transition rate on the photon energy of probe laser in a prolate ellipsoidal quantum dot with the ellipsoidal aspect ratio $\chi = 3$ (corresponding to the length of the semi-major axis $c = 75$ Å) in two cases: in the absence of the pump laser (dashed line) and in the presence of the pump laser (solid lines) with the detuning $\hbar\Delta\omega = 0$ meV.

The existence of the two distinct peaks in the absorption spectra of excitons under the effect of a strong pump laser resonating with the energy distance between the two quantized levels of the electron can be explained in Fig. 4. From Fig. 4, we see that when the pump laser is active, the initial electron levels are split into sub-levels. For example, two new levels $E_{100}^{e+}$ and $E_{100}^{e-}$ are separated from the initial level $E_{100}^{e}$, and two sub-levels $E_{110}^{e+}$ and $E_{110}^{e-}$ are split from the initial level $E_{110}^{e}$, determined according to Eqs. (43) and (44). Thus, when we use the probe laser, we find two interband transitions from hole level $E_{100}^{h}$ to two sub-levels of electron $E_{100}^{e+}$ and $E_{100}^{e-}$, obeying the selection rules of the interband transitions in quantum dots. This means in the absorption spectrum of excitons there appear two new absorption peaks of excitons. These two peaks are located symmetrically on both sides of the original peak according to the energy conservation law.

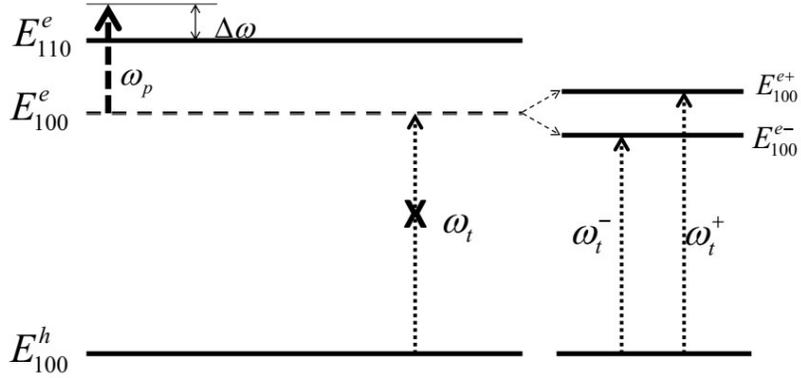

Fig. 4. The scheme of splitting of the energy levels of electrons under the effect of a strong pump laser resonant with two electron levels.

To better reveal the strong effect of the pump laser detuning on the absorption spectrum of excitons, we investigated the dependence of the transition rate according to the photon energy of probe laser when the ellipsoidal aspect ratio $\chi = 3$ in the absence of of the pump laser (dotted line) and in the presence of the pump laser with different detunings $\hbar\Delta\omega = 0.1$ meV (thin solid line), $\hbar\Delta\omega = 0.3$ meV (thick solid line), and $\hbar\Delta\omega = 0.5$ meV (dashed line) as shown in Fig. 5. From Fig. 5, we can see that in the presence of the pump laser, all absorption spectra of excitons contain two absorption peaks of excitons in the absorption spectra. The heights of two absorption peaks of excitons are different and depend on the pump laser detuning. When the pump laser detuning changes, the height of two exciton absorption peaks will also change, the more the detuning increases, the higher the height of one peak increases while the height of the other peak decreases. As the detuning increases, one absorption peak tends to move towards the original peak (which is the peak in the inactivity of the pump laser), the other is far from the position of the original peak, and this is clearly shown in Fig. 6. From Fig. 6, we also find that the more the detuning, the more redshifted two peaks. In addition, the detuning of the pump laser also strongly influences the transition rate of two absorption peaks as shown in Fig. 7.

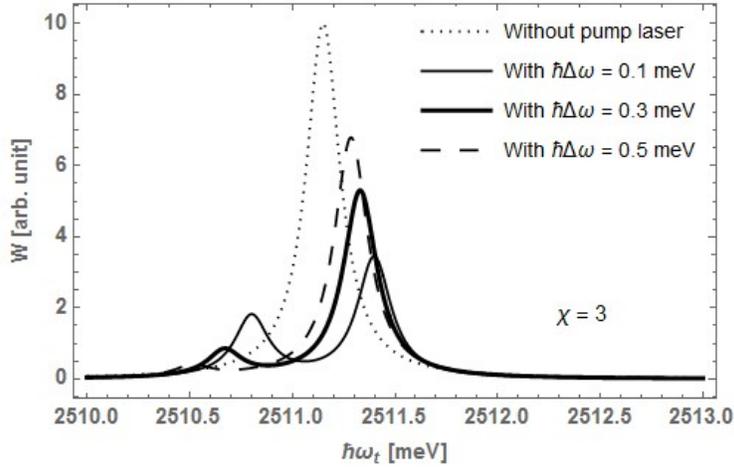

Fig. 5. The dependence of transition rate on the photon energy of probe laser in a prolate ellipsoidal quantum dot with the value of the ellipsoidal aspect ratio is $\chi = 3$ without the the pump laser effect (dotted line) and with the pump laser effect, with different detunings $\hbar\Delta\omega = 0.1$ meV (thin solid line), and $\hbar\Delta\omega = 0.3$ meV (thick solid line), and $\hbar\Delta\omega = 0.5$ meV (dashed line).

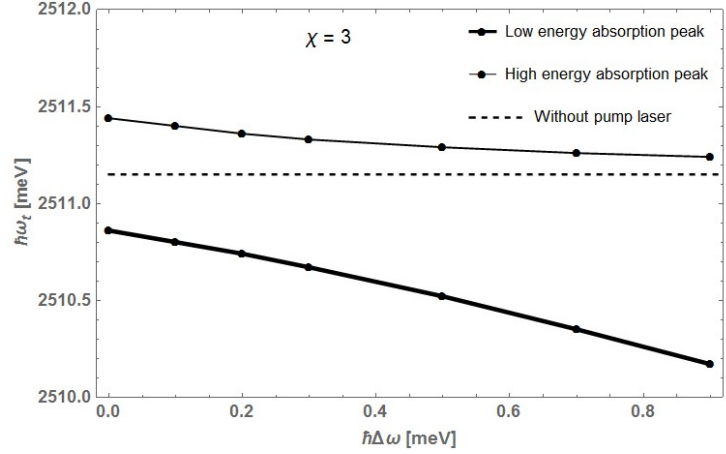

Fig. 6. The dependence of the photon energy of probe laser on the pump laser detuning of low-energy absorption peak (thick solid line), high-energy absorption peak (thin solid line), and original absorption peak in the absence of the pump laser (dashed line) in prolate ellipsoidal quantum dots, when $\chi = 3$ (corresponding to the length of the semi-major axis $c = 75$ Å).

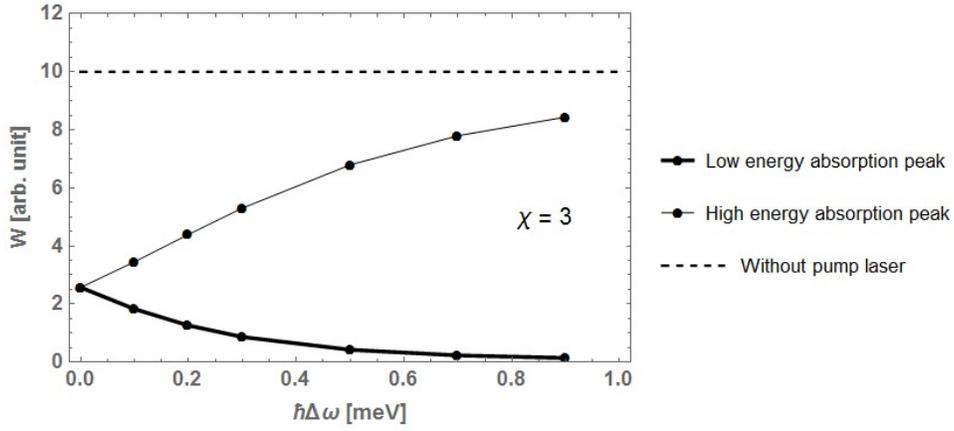

Fig. 7. The dependence of transition rate on the pump laser detuning of low-energy absorption peak (thick solid line), high-energy absorption peak (thin solid line), and original absorption peak in the absence of the pump laser (dashed line) in prolate ellipsoidal quantum dot when the ellipsoidal aspect ratio $\chi = 3$ (corresponding to the length of the semi-major axis $c = 75$ Å).

From Fig. 7, we see that when the pump laser detuning is increased, the transition rate of the high-energy absorption peak increases and approaches to the transition rate of original peak when the pump laser was not active, while the transition rate of low-energy absorption peak decreases and approaching to zero. At the same time, the transition rate of these two absorption peaks depends monotonously on the pump laser detuning.

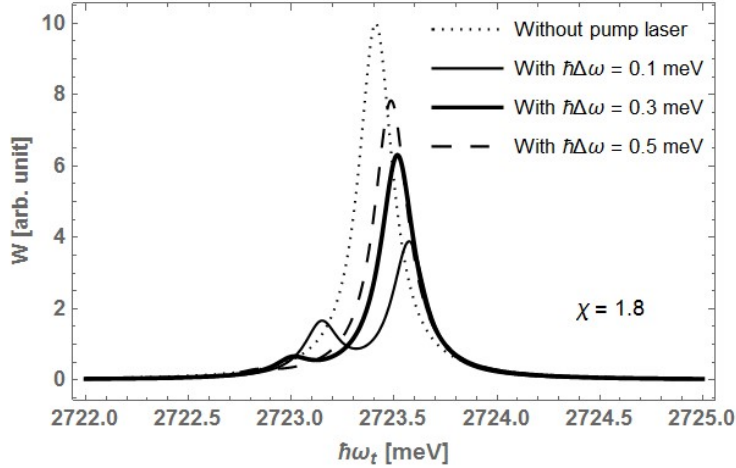

Fig. 8. The dependence of transition rate on the photon energy of probe laser in a prolate ellipsoidal quantum dot with the ellipsoidal aspect ratio $\chi = 1.8$ (corresponding to the length of the semi-major axis of $c = 45$ Å) in the absence of the pump laser (dotted line) and in the presence of the pump laser with different detunings $\hbar\Delta\omega = 0.1$ meV (thin solid line), and $\hbar\Delta\omega = 0.3$ meV (thick solid line) and $\hbar\Delta\omega = 0.5$ meV (dashed line).

Similar to Fig. 5, we also examine the dependence of the transition rate according to the photon energy of probe laser when $\chi = 1.8$ as shown in Fig. 8. Comparing Fig. 5 with Fig. 8, we found that the height of the low-energy absorption peak depends very sensitive on the value of $\chi$. With the same $\hbar\Delta\omega = 0.3$ meV value, when $\chi = 3$, the height of the low-energy absorption peak is still clearly observed; but with $\chi = 1.8$, the height of the low-energy absorption peak is very small. As the value of $\chi$ decreases, the low-energy absorption peak will almost disappear, and the high-energy absorption peak will reach the original peak. In addition, the location of these two exciton absorption peaks is also strongly dependent on the amplitude of pump laser as shown in the Fig. 9. In addition, in the case $\hbar\Delta\omega > 0$ as presented in Figs. 5 and 8, after being split (and decreasing peak intensity), the second peak intensity increases with the detuning and tends to return as in the case of no pump laser. This phenomenon can be explained as follows. When the detuning increases, the pump laser is hard to make an intraband transition between two electron levels (indicated by thick dashed arrow in Fig. 4), that means it is hard to connect those two electron levels to become one big degeneracy electron level. Hence, under effect of strong electric field of the pump laser, the symmetry of the initial electron levels (as well as the symmetry of this big degeneracy electron level) is hard to be broken and, hence, the initial electron levels are hard to be split into sub levels. This is the reason why we see one large splitting peak near the original peak (in the case of no pump laser), while another splitting peak nearly disappears. Therefore, when we probe the exciton spectrum, the second peak intensity increases with the detuning and tends to return as in the case of no pump laser and if the detuning is large enough, we will see only one peak in the spectrum.

Fig. 9 illustrates the absorption spectra of excitons in a prolate ellipsoidal quantum dot in the absence of the pump laser (dotted line) and in the presence of the pump laser with different amplitude of pump laser $A_p = 3\times 10^4$ V/cm (dashed line); $A_p = 6\times 10^4$ V/cm (thin solid line), $A_p = 8\times 10^4$ V/cm (thick solid line) when $\chi = 1.8$ and the pump laser detuning $\hbar\Delta\omega = 0$ meV. From Fig. 9, we find that when increasing the amplitude of the pump laser, the distance between the two exciton absorption peaks also increases, but these two peaks are located symmetrically on both sides of the original peak according to energy conservation law. In other words, the optical Stark effect can be controlled using the strong pump laser.

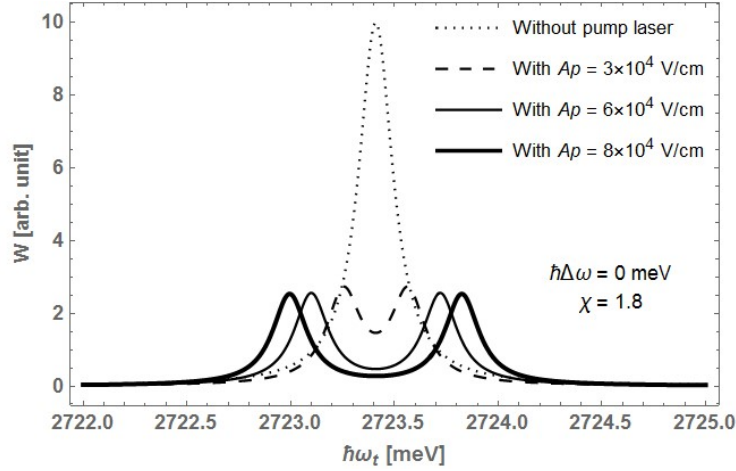

Fig. 9. The dependence of transition rate on the photon energy of probe laser in a prolate ellipsoidal quantum dot in the absence of the pump laser (dotted line) and in the presence of the pump laser with different amplitude of pump laser $A_p = 3 \times 10^4$ V/cm (dashed line), $A_p = 6 \times 10^4$ V/cm (thin solid line), $A_p = 8 \times 10^4$ V/cm (thick solid line) when $\chi = 1.8$ and the pump laser detuning $\hbar \Delta \omega = 0$ meV.

Next, we examine the effect of the $\chi$ value on the exciton absorption spectra. Fig. 10 describes the dependence of transition rate on the photon energy of probe laser with the different ellipsoidal aspect ratios $\chi = 3$ (dotted line), $\chi = 3.2$ (dashed line), and $\chi = 3.4$ (solid line) in the presence of the pump laser with detuning $\hbar \Delta \omega = 0.1$ meV. In all three cases, we can see two exciton absorption peaks, which again confirm the existence of the excitonic optical Stark effect in this structure. Besides, when $\chi$ was increased, both absorption peaks moved quickly to the lower energy region, corresponding to a red shift in the absorption spectrum. Therefore, changing the optical properties of prolate ellipsoidal quantum dot structures will become easier to control and more flexible by changing only two parameters semi-minor axis and semi-major axis of this quantum dot structure. This is one of the advantages of ellipsoidal quantum dot structures compared to spherical quantum dots.

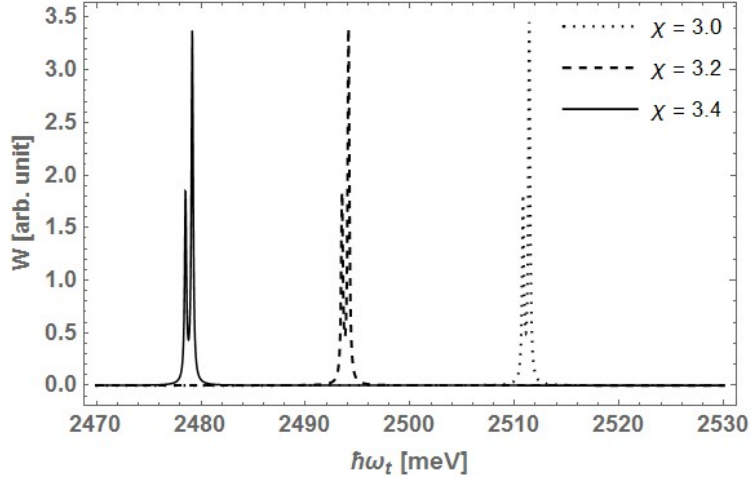

Fig. 10. The dependence of transition rate on the photon energy of probe laser with the different ellipsoidal aspect ratios $\chi = 3$ (or $c = 75$ Å) (dotted line), $\chi = 3.2$ (or $c = 80$ Å) (dashed line), and $\chi = 3.4$ (or $c = 85$ Å) (solid line) in the presence of the pump laser with detuning $\hbar\Delta\omega = 0.1$ meV.

To further investigate the properties of absorption spectra of interband transitions in prolate ellipsoidal quantum dots, we continue examining the shift of the photon energy of probe laser as a function of the ellipsoidal aspect ratio $\chi$ in the case of the pump laser is not switched on (Fig. 11). As shown in Fig. 11, we see a red shift in the absorption spectra when increasing the ellipsoidal aspect ratio $\chi$ as mentioned in Figure 10. In particular, the strongest shift occurs in the range of $1.2 < \chi < 3$ and weaker displacement appears when $\chi > 3$. This clearly shows the characteristic quantum size effect of quantum dot structures in general, and of prolate ellipsoidal quantum dots in particular. We think the result will be very useful for experimental studies in choosing the wavelength of probe laser to match the size of the quantum dots to study the existence of the optical Stark effect.

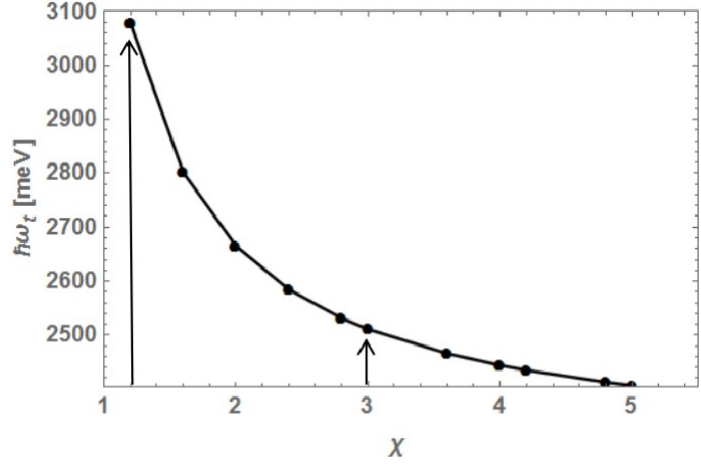

Fig. 11. The dependence of the photon energy of probe laser on the ellipsoidal aspect ratio $\chi$ in the absence of the pump laser.

Lastly, we compare the exciton absorption spectra in the prolate ellipsoidal quantum dot (Fig. 12a) and the spherical quantum dot (Fig. 12b) [11] with the same volume. Derived from Eq. (15), we consider the prolate ellipsoidal quantum dot in case $a = 25\,\text{Å}$ and $\chi = 4.096$, which show the same volume as the spherical quantum dot with radius $R = 40$ Å. From Fig. 12, we can see that, although two structures have the same volume, the exciton absorption spectra in two quantum dots of different shapes is completely different. This means that the exciton absorption spectra, or in other words, the optical Stark effect of exciton, not only depends on the pump laser detuning, size of quantum dots but also very sensitive on geometric shapes of quantum dots. The reason for the difference between the exciton absorption spectra of two quantum dot structures can be explained as follows. The shape of the quantum dots as well-known strongly affects the wavefunctions and the energy spectra of particles. Hence, the wavefunctions and the energy spectra of particles in quantum dots of different shapes are really distinct from others, though the quantum dots have the same volume. It leads to a change of the electronic transitions in those structures, resulting in a variation in the exciton absorption spectra. Moreover, because the renormalized wavefunction method is a quantum mechanical based approach, so our theory can be easily applied for the other low-dimensional structures as long as we can determine energy levels and wavefunctions of particles. In fact, we have applied our theory to the similar problems in QDs of various shapes as well as in quantum wires and quantum wells [11,12,37,38].

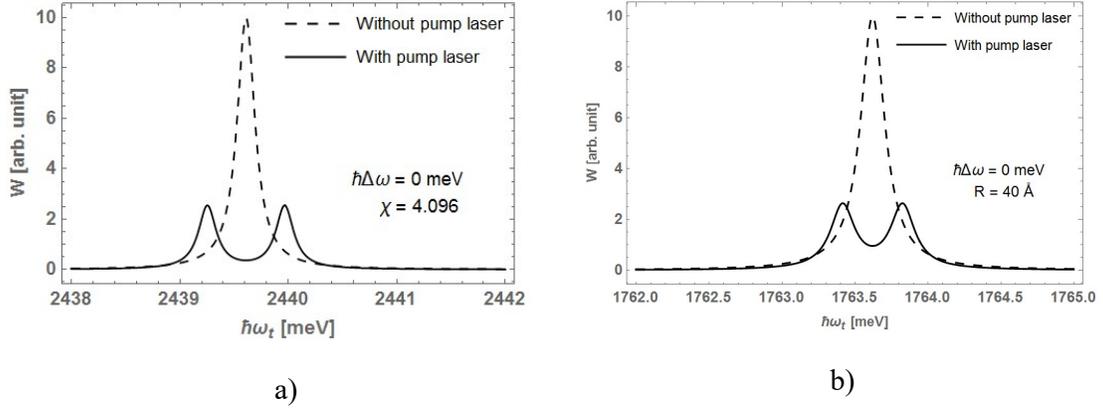

a)                                    b)

Fig. 12. The dependence of exciton absorption spectra on the photon energy of probe laser in the prolate ellipsoidal quantum dot (figure a) with the length of the semi-minor axis a = 25 Å, $\chi = 4.096$ (corresponding to the length of the semi- major axis $c = 102.4$ Å) and spherical quantum dot (figure b) with radius $R = 40$ Å with the same volume.

## 4. Conclusion

To sum up, in this study, we have investigated the characteristics of the exciton absorption spectra in the InGaAs/InAlAs prolate ellipsoidal quantum dot through calculating the transition rate of the interband transition using the renormalized wavefunction theory. The results show that when the two quantized energy levels of electrons are coupled by a polarized pump field, one initial absorption peak of exciton is split into distinct absorption peaks as an evidence of the existence of the three-level optical Stark effect. This is the consequence of the splitting of the electron quantized levels due to the effect of the strong resonant pump laser. We also find analytical expressions for the electron splitting levels and attempt to explain the mechanism of effect generation. Another important result is that the exciton absorption spectra are not only strongly affected by the energy of the pump laser but also depend strongly on geometric shapes of quantum dots. Specifically, with the same volume, the exciton absorption spectra in the spherical quantum dots and the prolate ellipsoidal quantum dots are completely different. Furthermore, when we increase the ellipsoidal aspect ratio of the prolate ellipsoidal quantum dots, a clear red shift in the optical absorption spectra was observed. The existence of two parameters of semi-minor axis and semi-major axis in the prolate ellipsoidal quantum dot structures makes the ability to control optical properties of the ellipsoidal quantum dots easier and more flexible than the spherical quantum dots. This is one of the advantages of the ellipsoidal quantum dot structures compared to the spherical quantum dots. We believe that the interesting features in the optical absorption spectra in the prolate ellipsoidal quantum dots when the optical Stark effect occurs have the great potential application to the development of computing devices and quantum information.

**Acknowledgment**

This research is funded by Vietnam Ministry of Education and Training (MOET) under grant number B-2020-DHH-06.

*E-mail: ltnbao@hueuni.edu.vn
**E-mail: dnthao@hueuni.edu.vn